\documentclass[11pt,letterpaper]{article}

\usepackage{jheppub}
\usepackage{color}
\usepackage{graphicx}
\usepackage{wrapfig,enumerate,slashed}

\usepackage{footmisc}
\usepackage{amsmath}
\usepackage{wasysym} 
\usepackage{graphicx}
\usepackage{color}
\usepackage{comment}
\usepackage{hyperref}

\newcommand{\be}{\begin{eqnarray}}
\newcommand{\ee}{\end{eqnarray}}

\newcommand{\bee}{\begin{eqnarray}}
\newcommand{\eee}{\end{eqnarray}}
\newcommand{\beeq}{\begin{equation}}
\newcommand{\eeeq}{\end{equation}}

\DeclareRobustCommand{\Sec}[1]{Sec.~\ref{#1}}

\DeclareRobustCommand{\Fig}[1]{Fig.~\ref{#1}}

\DeclareRobustCommand{\Eq}[1]{Eq.~(\ref{#1})}

\newcommand{\SU}{\text{SU}}

\newcommand{\U}{\text{U}}

\begin{document}

\title{S-Channel Dark Matter Simplified Models and Unitarity}

\author[a]{Christoph Englert,}
\author[b]{Matthew McCullough,}
\author[c]{and Michael Spannowsky}

\affiliation[a]{SUPA, School of Physics and Astronomy, University of
  Glasgow, Glasgow G12 8QQ, UK}
\affiliation[b]{Theory Division, CERN, 1211 Geneva 23, Switzerland}
\affiliation[c]{Institute for Particle Physics Phenomenology, Department of Physics, Durham University, DH1 3LE, UK}

\emailAdd{christoph.englert@cern.ch}
\emailAdd{matthew.mccullough@cern.ch}
\emailAdd{michael.spannowsky@durham.ac.uk}

\abstract{The ultraviolet structure of $s$-channel mediator dark matter simplified models at hadron colliders is considered.  In terms of commonly studied $s$-channel mediator simplified models it is argued that at arbitrarily high energies the perturbative description of dark matter production in high energy scattering at hadron colliders will break down in a number of cases.   This is analogous to the well documented breakdown of an EFT description of dark matter collider production.  With this in mind, to diagnose whether or not the use of simplified models at the LHC is valid, perturbative unitarity of the scattering amplitude in the processes relevant to LHC dark matter searches is studied.  The results are as one would expect: at the LHC and future proton colliders the simplified model descriptions of dark matter production are in general valid.  As a result of the general discussion, a simple new class of previously unconsidered `Fermiophobic Scalar'  simplified models is proposed, in which a scalar mediator couples to electroweak vector bosons.  The Fermiophobic simplified model is well motivated and exhibits interesting collider and direct detection phenomenology.}

\preprint{DCPT/16/52, IPPP/16/26, CERN-TH-2016-094}

\maketitle


\section{Introduction}
\label{sec:intro}
Proposals to search for dark matter through missing energy signatures at colliders have been considered for a long time.  It is no surprise then that the LHC has begun to play an increasingly prominent role in the multifaceted search for particle dark matter.  Only experiments sensitive to the cosmological abundance of dark matter, such as direct and indirect detection experiments, can unequivocally provide evidence for the dark matter particles dominating the matter abundance of the Universe.  However, LHC searches for tell-tale missing energy signatures can provide a powerful complementary probe by searching for evidence of neutral particles that are stable, at the least, on collider timescales.  In combination, the cornucopia of different dark matter probes could allow in principle for a detailed characterisation of dark matter interactions at high energies.

In order to search for dark matter at the LHC it is necessary to compare observed processes involving missing energy against both background and signal predictions.  Backgrounds may be estimated in a number of ways, often by combining SM theory with data driven techniques, such as using $Z\to\mu^+ \mu^-$ observations to model the $Z\to\nu \nu$ contribution to missing energy signatures.  However, for signal predictions we require a microscopic description to model missing energy signatures.  The number of potential UV-complete models is enormous, leading to an already large literature on the subject, and thus it is not feasible to study every dark matter model independently.  This has led to the consideration of effective theory approaches to DM interactions \cite{Beltran:2008xg,Cao:2009uw,Beltran:2010ww,Goodman:2010yf,Bai:2010hh,Goodman:2010ku,Goodman:2010qn,Rajaraman:2011wf,Fox:2011pm}.  In certain circumstances the interpretation of a measurement using an effective theory approach may become invalid (see Sec.~5 of \cite{Fox:2011pm} for a discussion, and in more detail \cite{Fox:2011fx,Shoemaker:2011vi,Fox:2012ee,Busoni:2013lha,Buchmueller:2013dya,Busoni:2014sya,Endo:2014mja,Busoni:2014haa,Racco:2015dxa,Bell:2015sza}).  As a result, the use of simplified models \cite{Dudas:2009uq,Fox:2011qd,Goodman:2011jq,An:2012va,Frandsen:2012rk,Dreiner:2013vla,Cotta:2013jna,Alves:2013tqa,Arcadi:2013qia,deSimone:2014pda,Garny:2014waa,Lebedev:2014bba,Fairbairn:2014aqa,Buchmueller:2014yoa,Abdallah:2014hon,Duerr:2014wra,Buckley:2014fba,Busoni:2014gta,Harris:2014hga,Alves:2015pea,Jacques:2015zha,Martin-Lozano:2015vva,Haisch:2015ioa,Xiang:2015lfa,Chala:2015ama,Blennow:2015gta,Godbole:2015gma,Abdallah:2015ter,Khoze:2015sra,Duerr:2015wfa,Alves:2015mua,Abercrombie:2015wmb,Alves:2015dya,Choudhury:2015lha,Heisig:2015ira,Backovic:2015soa,Harris:2015kda,Kahlhoefer:2015bea,Bell:2015rdw,Haisch:2016usn,Brennan:2016xjh,Boveia:2016mrp,Jackson:2013pjq,Jackson:2013rqp,Duerr:2013lka,LopezHonorez:2012kv,Basso:2011na,Baek:2012uj,Esch:2013rta,Freitas:2015hsa,Chun:2010ve,Brooke:2016vlw}, has increasingly become a central component of collider dark matter searches.  Most early examples of simplified models were motivated by the consideration of specific scenarios, however more recently they have become motivated by the desire for greater generality in LHC searches and also to address the shortcomings of the effective theory approach whenever they arise.  The purpose of the simplified model approach is to characterise the dark matter production processes present in UV-complete models without having to specify the entire UV-completion (see e.g.\ \cite{Abdallah:2015ter} for a discussion).  These simplified models often contain both dark matter and mediator particles, where the latter provide the link between the visible SM particles and dark matter.

As stated, by design these simplified models do not contain all of the ingredients present in a UV-complete model of dark matter.  In \Sec{sec:models} we will argue, in an approach aimed at a non-expert reader, that in some cases this essentially means that in some scattering processes at high energy the simplified model description will break down.  This breakdown may manifest as a violation of unitarity.  While this was briefly suggested for axial-vector mediators in \cite{Chala:2015ama}, it was explicitly demonstrated and systematically investigated in \cite{Kahlhoefer:2015bea,Haisch:2016usn}.\footnote{Perturbative unitarity has played an important role in physics beyond the standard model for some time, including for dark matter \cite{Lee:1977eg,Chanowitz:1978mv,Griest:1989wd,Gunion:1990kf,Walker:2013hka,Hedri:2014mua,Englert:2015oga}.}  This breakdown is often due to the fact that, although a simplified model may contain apparently renormalisable operators, in many cases in a fully electroweak gauge invariant theory these operators must arise as a higher dimension operator or else they signal the need for additional particles or couplings in order to respect gauge invariance \cite{Bell:2015sza,Chala:2015ama,Kahlhoefer:2015bea,Haisch:2016usn}.

Taking the basic arguments of \Sec{sec:models} as guidance, in \Sec{sec:newmodel} we propose an additional class of `fermiophobic' simplified models coupling a scalar mediator to vector bosons, which to our knowledge has not yet been considered in the literature.\footnote{The recent work of \cite{Brooke:2016vlw} considered the Higgs mediator scenario and a new model with DM directly coupled to SM gauge bosons, however a new (non-Higgs boson) scalar mediator coupled to gauge bosons was not considered.}  This model is as relevant for dark matter phenomenology as other commonly considered scalar mediator models, including scenarios which couple a scalar mediator to quarks.  The fermiophobic model is interesting as the strongest signals may emerge in MET+VBF channels, rather than monojets, and direct detection is also greatly suppressed.

Since, as argued, some simplified models will break down in particular scattering processes at high energies, we ask whether this implies that the simplified model program should be modified?  To answer this question we take a pragmatic approach motivated by the spirit of simplified models.  The intended purpose of a simplified model is not as a UV-complete theory of dark matter interactions (which would hopefully be fully unitary in every respect), but is rather a tool by which to model the missing energy signatures at colliders.  Based on this we consider the most pertinent question with regard to unitarity of scattering amplitudes in the context of dark matter simplified models to be:  Is unitarity violated in the \emph{specific dark matter production process under study}?  The reason is that if unitarity is violated in other processes not relevant to missing energy collider signatures then it may be restored by additional particles or fields which may or may not significantly impact the collider phenomenology.  Whether these new states or couplings will modify the dark matter phenomenology becomes a model-dependent issue, clearly discussed in \cite{Kahlhoefer:2015bea}.  However, if unitarity is violated in the specific collider process under consideration then additional particles and couplings will certainly be required to restore unitarity in that specific process and the dark matter collider phenomenology must be altered.

To this end, in \Sec{sec:unitarity} we study the dominant $2\to2$ mediator production processes in popular classes of $s$-channel mediator simplified models, as well as the model proposed in \Sec{sec:newmodel}.  For each process we determine the coupling at which unitarity is violated at specific center of mass energies.  In every case this coupling is so large that unitarity violation is simply a result of the breakdown of perturbation theory, rather than a structural breakdown of the simplified model description itself. We offer conclusions in \Sec{sec:conclusions}.

\section{Modelling Dark Matter Production at the LHC}
\label{sec:models}
Now we present a broad discussion which aims to sketch potential issues in the ultraviolet behaviour of some dark matter simplified models, providing examples where necessary.  As already discussed, in order to search for dark matter at the LHC it is necessary to have some underlying microscopic description of dark matter interactions (a Lagrangian) in order to predict dark matter production signatures at the LHC.  A first step towards modelling dark matter interactions would be to add only a dark matter species (let us call it $\chi$) to the Standard Model particles.  In more involved models we may also wish to add other particles, such as a particle (let us call it $\phi$) which mediates interactions between the Standard Model and the dark sector.  This will capture a richer phenomenology.
\begin{figure}[!t]
\begin{centering}
\includegraphics[width=0.6\textwidth]{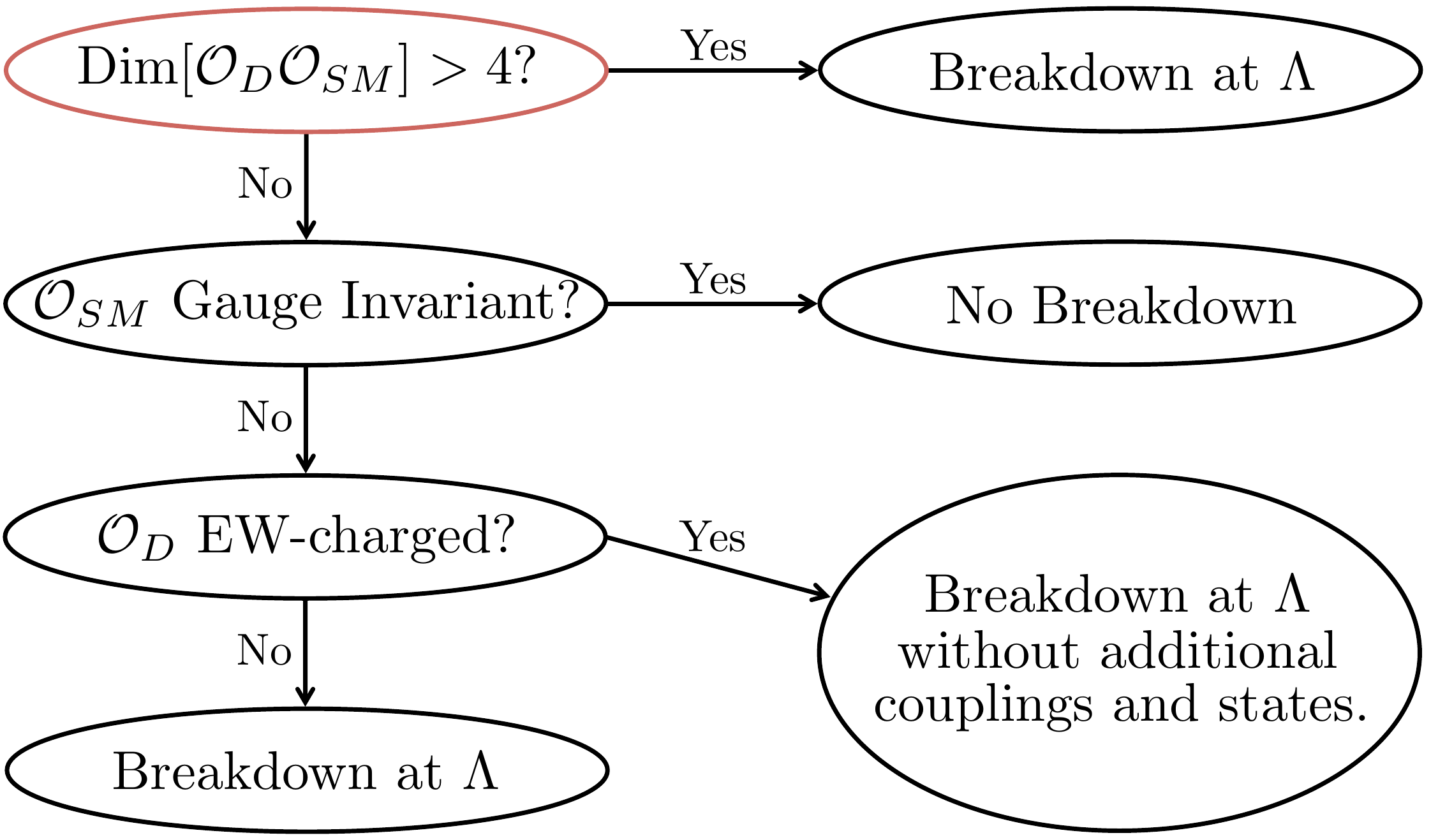}
\caption{A flowchart, starting with the red oval, detailing various possibilities for the structure of a simplified model for dark matter interactions at colliders.  In all cases $\Lambda$ is used as a generic, model dependent, high energy scale.}
\label{fig:flow}
\end{centering}
\end{figure}
To model the interactions of these particles with the SM particles we will construct a Lagrangian containing various operators.  These operators contain the relevant fields and may also involve spacetime derivatives, Dirac matrices, and so on.  The operators should also respect any symmetries, such as the SM gauge symmetries, or a DM stabilization symmetry.  We will write the operator containing SM fields as $\mathcal{O}_{SM}$ and the one containing dark sector states, such as the DM or a mediator, as $\mathcal{O}_D$.

More generally, we may use this recipe involving operators to consider broad classes of interactions.  Henceforth, we will assume that $\mathcal{O}_D$ is completely neutral under QED and QCD.  This assumption is very well motivated if $\mathcal{O}_D$ contains the DM, however if $\mathcal{O}_D$ contains the mediator particle then this is essentially implying that the mediator is not charged and does not carry colour, thus the discussion contained here will not cover many t-channel mediator models, which themselves have very interesting phenomenology \cite{Busoni:2014haa,Chang:2013oia,An:2013xka,Bai:2013iqa,DiFranzo:2013vra,Papucci:2014iwa,Agrawal:2011ze,Kile:2013ola,Agrawal:2014una}.   However it does cover the $s$-channel class of models.  Under these assumptions we see that whatever the combination of fields in $\mathcal{O}_{SM}$, the total charge and total colour charge of $\mathcal{O}_{SM}$ must be zero.

\subsection{Is Dim$[\mathcal{O}_D \mathcal{O}_{SM}] > 4$?}
Let us now discuss the energy scales at which we will employ the prescription described above.  If we wish to consider probing the dark sector at very low energies, such as in direct detection, then the theory need only be valid up to relatively low energy scales.  In practice, validity up to energies of a few MeV should certainly suffice for direct detection. However, if we wish to study the dark sector at higher energies, in particular at the LHC, then the model must be valid at LHC energy scales.  This is an unavoidable constraint.  Broadly speaking, if the operator is not renormalizable, which in practise means that the operator dimension satisfies Dim$[\mathcal{O}_D \mathcal{O}_{SM}] = n > 4$, then we must write the relevant term in the Lagrangian as $\mathcal{L} = \lambda \mathcal{O}_D \mathcal{O}_{SM}/\Lambda^{n-4}$, where $\Lambda$ is the typical scale at which this description will breakdown and cease to be appropriate.   If this operator is being probed at the LHC at a comparable energy $E \sim \Lambda$ then this description should not be used.  There are a number of reasons why the validity of the theory is breaking down.  First, the theory is no longer effectively capturing the effects of the unknown physics as other higher dimension operators that have not been included are likely to be equally, or even more, relevant.  Second, other problems may arise, such as the loss of unitarity in the scattering amplitudes.  Essentially, the full microscopic theory begins to be resolved in scattering processes and it is necessary to employ the full microscopic theory to reliably predict cross sections.  This potential issue in modelling dark matter production at the LHC is the first step shown in \Fig{fig:flow}.

Let us consider two simple examples to illustrate the point.  In the first, we may include a new contribution to the Lagrangian which is the renormalizable Higgs portal interaction for a scalar DM particle: $\mathcal{L} = \lambda \chi^2 |H|^2$.  In this case we have $\mathcal{L} = \lambda \mathcal{O}_D \mathcal{O}_{SM}$, where $\mathcal{O}_D = \chi^2$, $\mathcal{O}_{SM} = |H|^2$ and the theory will remain classically valid up to high energies.  For a second example, we could write a nonrenormalizable operator $\mathcal{L} =  \mathcal{O}_D \mathcal{O}_{SM}/\Lambda$, where again $\mathcal{O}_D = \chi^2$ and now $\mathcal{O}_{SM} = \overline{q} q$ is a SM quark pair combination.  As this theory is nonrenormalizable it will only be valid up to energy scales $E\sim \Lambda$.

It should be noted that it is still possible to set limits in this case, by truncating the energies considered as suggested in \cite{Racco:2015dxa}.  Furthermore, perturbative unitarity constraints may also be considered to understand the energies at which the effective theory breaks down \cite{Endo:2014mja,Racco:2015dxa}.

\subsection{Is $\mathcal{O}_{SM}$ Gauge Invariant?}
If the operator $\mathcal{O}_{SM}$ is invariant under QED and QCD, and satisfies Dim$[\mathcal{O}_D \mathcal{O}_{SM}] \leq 4$, is it safe to assume that the theory is valid at LHC energy scales?  It is commonly assumed within the context of simplified dark matter models that this is true, however we will see that this is not necessarily the case.  
The reason is that the LHC operates at energies at, or above, the weak scale.  As a result the full EW gauge symmetry must be considered in addition to QED and QCD.  This section covers the second step in \Fig{fig:flow}.

Let us consider the most innocuous possibility, where Dim$[\mathcal{O}_D \mathcal{O}_{SM}] \leq 4$ and $\mathcal{O}_{SM}$ is invariant under the full SM gauge group $\SU(3)_C \times \SU(2)_L \times \U(1)_Y$.  In this case the theory is renormalizable and gauge invariant thus it will not break down at the classical level at LHC energies.\footnote{We are intentionally not considering quantum effects which may cause a description to breakdown, such as RG running into strong coupling, or anomalous violation of a classical symmetry.}  Theories of this class are clearly very attractive, however this option is unfortunately too limiting.   Well known examples of fully gauge-invariant couplings include the Higgs portal and hypercharge kinetic mixing scenarios, as well as vector mediators corresponding to new $\text{U}(1)$ gauge forces.  If we wish to consider a wider range of possibilities it is necessary to consider the case where $\mathcal{O}_{SM}$ is not invariant under the full EW gauge group.

\subsection{Is $\mathcal{O}_{D}$ EW-charged?}
If $\mathcal{O}_{SM}$ is not invariant under the full EW gauge group we are lead to a bifurcation in the logical possibilities for a DM model.  The two possibilities are
\begin{itemize}
\item a) $\mathcal{O}_{D}$ is also charged under the EW gauge group and is essentially the conjugate of $\mathcal{O}_{SM}$, such that the combination $\mathcal{O}_D \mathcal{O}_{SM}$ is gauge invariant.
\item b) $\mathcal{O}_{D}$ is completely uncharged under the EW gauge group. This implies that the operator $\mathcal{O}_{SM}$ initially arose as a gauge-invariant operator $\mathcal{O}_H \mathcal{O}'_{SM}$ where in the EW-breaking vacuum $\langle H \rangle \rightarrow v_{EW}$ and $\mathcal{O}_H\rightarrow$ constant.\footnote{Of course it could also be the case that $\mathcal{O}_{D}$ is charged under the EW gauge group, but is not the conjugate of $\mathcal{O}'_{SM}$.  In this instance you also need an additional operator $\mathcal{O}_H$ and the conclusions of both a) and b) will apply.}
\end{itemize}
We will now consider each possibility in turn.  The outcome is shown in the final step of \Fig{fig:flow}.

a)  If we assume that $\mathcal{O}_{D}$ is also charged under the full EW gauge group then there must exist new charged states.  The maximum mass splitting between states contained within an $\SU(2)_L$ multiplet is in general not large since it can only come from electroweak symmetry breaking.  Any mass splitting should be at most $\mathcal{O}(\text{few}\times v_{EW})$.  Ignoring these extra charged states corresponds to assuming they are decoupled from the theory.  For this to be the case very large couplings between these extra states and the SM Higgs are required.  Similarly, an EW charge for $\mathcal{O}_{D}$ also implies couplings of the states in $\mathcal{O}_{D}$ to the EW gauge bosons.  The absence of the extra states and the EW couplings will generally lead to a breakdown of unitarity at some scale $\Lambda$.  A clear early treatment of the required couplings to EW gauge bosons was presented some time ago in \cite{Gunion:1990kf}.

b)  If $\mathcal{O}_{D}$ is completely uncharged under the EW gauge group then, as described in the bullet points, $\mathcal{O}_{SM}$ must have initially come from a gauge-invariant operator of the form $\mathcal{O}'_{SM} \mathcal{O}_H/\Lambda_H^{n_H}$ which became $\mathcal{O}_H \mathcal{O}'_{SM} \rightarrow \mathcal{O}_{SM}$ when the vacuum expectation value for the Higgs is included.  Thus it must have come from a higher dimension operator combination $\mathcal{O}_{D} \mathcal{O}'_{SM} \mathcal{O}_H/\Lambda_H^{n_H}$ and once again this description will break down at energies of $\Lambda_H$, where $\Lambda_H$ may be at energies probed by the LHC.

\subsection{Checking the validity of the description.}
While UV-complete models are certainly amongst the best motivated scenarios to search for, restricting to only UV-complete models is not pragmatic as it would be seemingly excessive to construct numerous different UV-complete models, including DM and messenger fields with varieties of EW gauge charges and couplings, simply to study the tell-tale missing energy signatures at the LHC.  Thus the only reasonable choice seems to be the continued use of simplified models.  However we have just argued that in many cases they break down at high energies unless new states or couplings are introduced.  In this work, to address these potential inconsistencies, we will take an approach inspired by the simplified models paradigm.  

Essentially, the most immediate constraint to impose is that the simplified model description does not break down in the specific process being studied at the LHC.  In \Sec{sec:unit} we will study a variety of simplified models and check, on a case-by-case basis, whether perturbative unitarity is violated at the level of the scattering amplitude in the specific DM searches performed at the LHC.  This is a necessary condition for the validity of the simplified model description of dark matter production.

\section{A New Fermiophobic Scalar Mediator Simplified Model}\label{sec:newmodel}
Let us consider the commonly studied scalar mediator model: $\lambda_q \phi \overline{q} q$.  This coupling explicitly violates electroweak gauge invariance, thus it can only be embedded in a gauge invariant structure at the weak scale if additional structure is invoked.  For example, it could arise as a higher dimension operator
\be
\mathcal{L} = \frac{\phi}{\Lambda} \lambda_q H Q U^c ~~,
\ee
or, in a renormalizable model, if $\phi$ belongs, at least in some admixture, to a multiplet with the same quantum numbers as the Higgs.  This latter possibility could arise if $\phi$ is the heavy Higgs of a two Higgs doublet model or if $\phi$ is a singlet scalar mixed with the Higgs.  Both possibilities predict new states or modified couplings that can also be searched for.

Since the coupling $\lambda_q \phi \overline{q} q$ violates gauge invariance, why do we not also consider the coupling
\be
\mathcal{L} = c_\phi \phi \left( \frac{M_W^2}{v} W^{+\mu} W^-_\mu + \frac{M_Z^2}{2 v} Z^{\mu} Z_\mu \right)~~?
\label{eq:vectorscalar}
\ee
In writing this we are violating gauge invariance in the same manner as with the couplings to quarks.  Let us now consider embedding this within a gauge invariant structure at the weak scale.  It could arise as a higher dimension operator, such as
\be
\mathcal{L} = \frac{\phi}{\Lambda} |D_\mu H|^2 \to  \frac{\phi}{\Lambda} \left( M_W^2 W^{+\mu} W^-_\mu + \frac{1}{2} M_Z^2 Z^{\mu} Z_\mu \right) ~~,
\ee
in exactly the same manner as the coupling of a scalar mediator to quarks, or in a renormalisable model if $\phi$ belongs, at least in some admixture, to a multiplet with the same quantum numbers as the Higgs.

\begin{figure}[!t]
\begin{centering}
\includegraphics[width=0.8\textwidth]{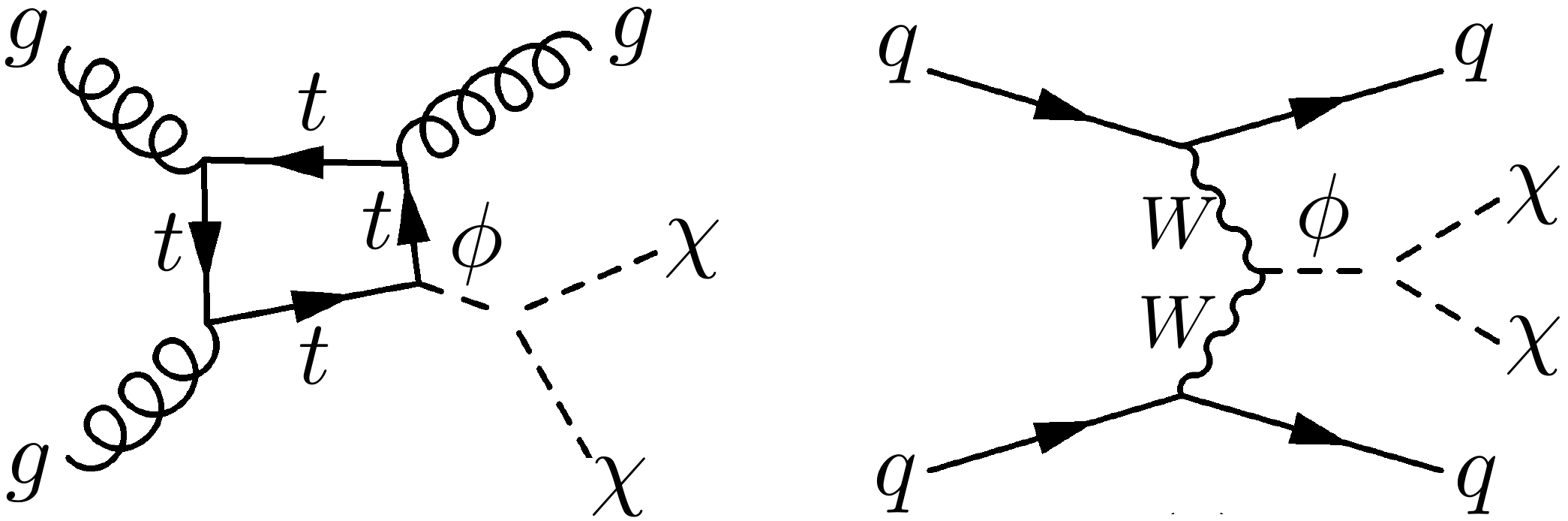}
\caption{Monojets in association with DM production at the LHC (left), which leads to weaker constraints than from VBF production (right) in simple Higgs portal models.}
\label{fig:VBF}
\end{centering}
\end{figure}

Let us consider a simple UV-complete model to illustrate these points and further investigate the nature of the couplings.  We will consider a scalar mediator coupled to dark matter as $\phi \chi^2$.  This scalar mediator can obtain couplings to SM states via a Higgs portal mixing with the Higgs.  Due to this mixing it inherits all of the SM Higgs couplings, suppressed by a factor $\sin \theta$, where $\theta$ is the mixing angle.  This model thus has couplings to quarks, leptons, and vector bosons
\be
\mathcal{L} = \sin \theta\, \phi \left( \frac{m_q}{v} \overline{q} q +  \frac{m_l}{v} \overline{l} l +2\left( \frac{M_W^2}{v} W^{+\mu} W^-_\mu + \frac{M_Z^2}{2 v} Z^{\mu} Z_\mu \right) \right) ~~.
\ee
First of all, this demonstrates that in UV-complete models realising the $\lambda_q \phi \overline{q} q$ interaction, the interaction of \Eq{eq:vectorscalar} also typically arises.  Second, the results of \cite{CMS-PAS-HIG-15-012,Craig:2014lda,Brooke:2016vlw} demonstrate that when a mediator has these couplings the strongest collider bounds will arise from VBF production of the DM, shown in \Fig{fig:VBF}.  Since the monojet bounds arise from the mediator couplings to quarks, and the VBF bounds from the mediator couplings to vectors, it is clear that it may be possible to overlook the strongest probes of DM for scalar mediators at the LHC if one only considers the $\lambda_q \phi \overline{q} q$ interaction for scalar mediators.

However, when considering non-DM observables, the situation is not as clear-cut.  As has been pointed out previously,\footnote{See \cite{Brooke:2016vlw} for a recent discussion.} in many UV-complete models that lead to the coupling in \Eq{eq:vectorscalar}, the Higgs couplings to vector bosons will also be modified.  There are already strong constraints on such modifications from precision electroweak measurements and also from LHC Higgs measurements, whereas constraints on modifications of the Higgs couplings to fermions are weaker.  Thus it has been argued that for the purposes of simplified scalar models, couplings to fermions are better motivated than \Eq{eq:vectorscalar}.  This point of view also has some theoretical support.  On one hand, the Higgs portal model shows that both Higgs-vector and Higgs-fermion couplings scale in the same way, making them equally well motivated.  However, an obvious counterexample is a type-II 2HDM.  If we take the usual parameterisation $\beta-\alpha\to \delta+\pi/2$, where $\delta$ is small for a SM-like Higgs, then the Higgs-vector coupling modification scales as $(1-\delta^2/2)$, whereas the Higgs-top coupling modification scales as $(1-\delta \cot \beta)$, which decouples much less rapidly.  Moreover, if the heavy Higgs of a type-II 2HDM is the mediator, then its coupling to vector bosons decouples as $\delta$, whereas its coupling to top quarks may persist as $-\cot \beta$ even when $\delta\to0$.  Thus there are clearly well motivated UV-complete scenarios in which a) a mediator coupling to vectors is as well motivated as a coupling to top quarks, and b) a mediator coupling mostly to top quarks is well motivated.

It is also worth mentioning that in scenarios with partial electroweak symmetry breaking from non-doublet $SU(2)_L$ gauge representations~\cite{Georgi:1985nv,Chanowitz:1985ug}, the tight correlation of electroweak precision observables with the VBF phenomenology can be relaxed \cite{Englert:2013zpa}. This is again directly related to new unitarity-related exotics in the particle spectrum~\cite{Grinstein:2013fia}.  Thus in even more exotic scenarios there can be significant modifications of the electroweak sector, involving new scalar mediator candidates with large couplings to vector bosons, that are consistent with current bounds.

In summary, it is clear that, as the relative importance of the two couplings is model-dependent, it is important to consider a complete set of couplings between the SM and mediators, with the aim of covering as many processes as possible to ensure that channels which may be the most constraining when limits are projected onto specific UV-complete models are not overlooked.  The fermiophobic model adds to this set of couplings and can be searched for in mono-boson+MET and VBF+MET searches.

\section{Unitarity constraints on dark matter production at hadron colliders}\label{sec:unitarity}
In order to investigate perturbative unitarity in $s$-channel mediator simplified models we will consider four simplified models which are commonly studied.  The models include scalar ($\phi \equiv S$), pseudoscalar ($\phi \equiv P$), vector ($\phi \equiv V$), and axial-vector ($\phi \equiv A$) mediators with the interactions
\begin{align}
\label{eq:LS} 
\mathcal{L}_{\mathrm{S}}&\supset\,  - \sum_q c_S \lambda_{h,q} S \, \bar{q}q -\,\frac{1}{2}m_{\rm MED}^2 S^2 + \mathcal{L}(S , \bar{\chi}, \chi ) \,,
 \\
 \label{eq:LP} 
\mathcal{L}_{\rm{P}}&\supset\,  -\sum_q  i c_P \lambda_{h,q} P \, \bar{q}  \gamma^5q -\,\frac{1}{2}m_{\rm MED}^2 P^2 + \mathcal{L}(P , \bar{\chi}, \chi ) \,,
 \\
 \label{eq:LV} 
\mathcal{L}_{\mathrm{V}}&\supset \, -\sum_q c_V V_{\mu} \bar{q}\gamma^{\mu}q - \frac{1}{2}m_{\rm MED}^2 V_{\mu} V^{\mu} + \mathcal{L}(V , \bar{\chi}, \chi ) \,,
\\
\label{eq:LA} 
\mathcal{L}_{\rm{A}}&\supset\, -\sum_q c_A A_{\mu} \bar{q}\gamma^{\mu}\gamma^5q - \frac{1}{2}m_{\rm MED}^2 A_{\mu} A^{\mu}+ \mathcal{L}(A , \bar{\chi}, \chi ) \,,
\end{align}
where we have assumed the scalar and pseudoscalar couplings are proportional to the Higgs Yukawa couplings, such that $c_S = 1$ corresponds to a scalar coupling to fermions which is identical to the Higgs boson coupling.  For the most part we will consider the case where the mediator may decay to dark matter $m_{\rm MED} > 2 m_{\rm DM}$, thus for our study the dark matter coupling and the nature of the dark matter particle, contained in $\mathcal{L}(P , \bar{\chi}, \chi )$, is largely irrelevant.  As long as the mediator decays into the invisible sector we may study unitarity in dark matter production processes by considering $2 \to 2$ scattering where the mediator is produced on shell.

In addition to the commonly studied simplified models shown above we will also consider the couplings proposed in \Sec{sec:newmodel} for a scalar mediator, again normalized to the coupling of the SM Higgs boson
\begin{equation}
\mathcal{L}_{\mathrm{S,VV}} \supset\,  -2 c_{S,VV} S \left(\frac{M_W^2}{v} W^{+\mu} W^-_\mu +\frac{M_Z^2}{2 v} Z^{\mu} Z_\mu  \right) \,  -\,\frac{1}{2}m_{\rm MED}^2 S^2 + \mathcal{L}(S , \bar{\chi}, \chi ) \,.
\label{eq:SVV} 
\end{equation}
where, as with the scalar coupling to quarks, $c_{S,VV}\to 1$ reproduces a mediator coupling equal to the usual Higgs boson coupling.  This completes the list of simplified models studied here.  In all cases we will assume a DM mass of $M_{DM}=50$ GeV and we will compute the mediator decay width into SM states and DM using the formul\ae ~of \cite{Boveia:2016mrp}.  Cases will also be shown with $M_\phi < 2 M_{DM}$, where the unitarity bounds from mediator production still apply in the same way, however dark matter production occurs in this case only through an off-shell mediator.

\subsection{Unitarity constraints on mediator production}\label{sec:unit}
\begin{figure}[!t]
\begin{centering}
\includegraphics[width=0.7\textwidth]{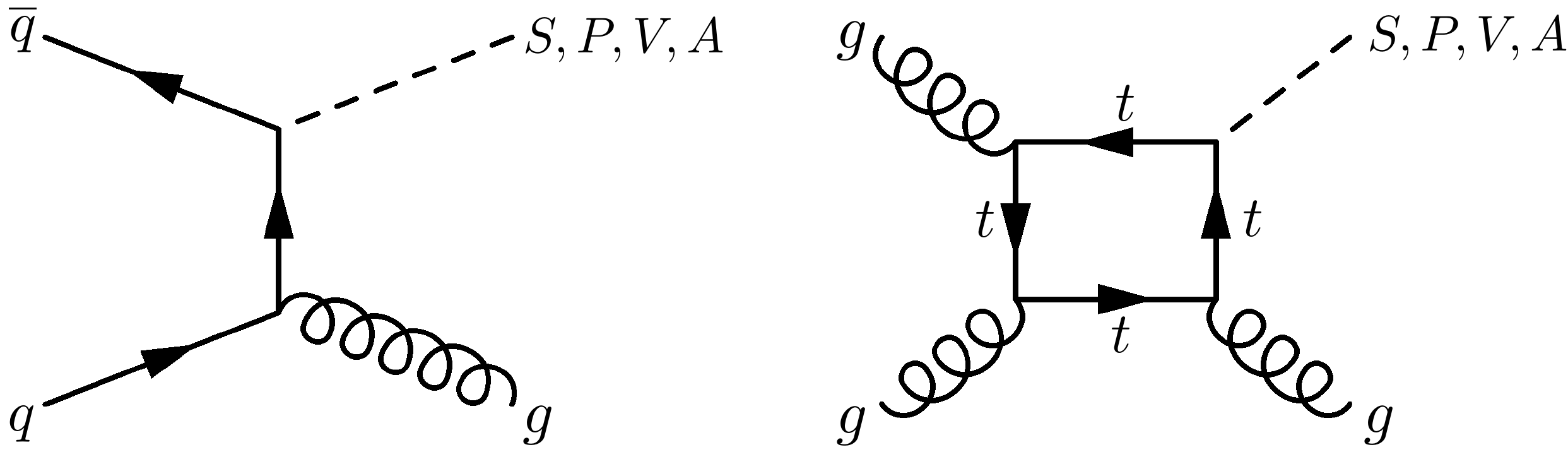}
\caption{Feynman diagrams for the relevant monojet production processes in commonly studied simplified models of $s$-channel mediators.}
\label{fig:monojet}
\end{centering}
\end{figure}

\begin{figure}[!t]
\begin{centering}
\includegraphics[width=0.7\textwidth]{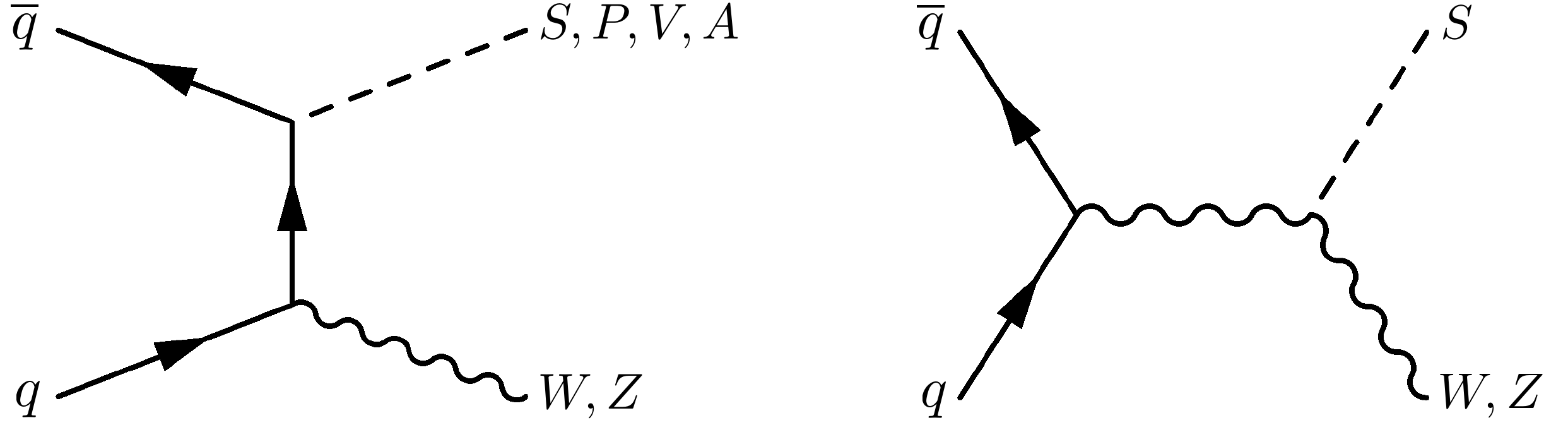}
\caption{Feynman diagram for mono-$W,Z$ processes, including mediator production whenever Higgs-like couplings of scalar mediators to massive vector bosons are also included (right).}
\label{fig:monoz}
\end{centering}
\end{figure}

We will study the mediator production process shown in \Fig{fig:monojet} and \Fig{fig:monoz}.  Throughout we will only consider the parton-level CM energy $\sqrt{\hat{s}}$.

We evaluate unitarity constraints with helicity amplitudes of {\tt MadGraph 5} \cite{Alwall:2011uj} using the {\tt UFO} interface \cite{Degrande:2011ua} provided by {\tt Feynrules} \cite{Alloul:2013bka} for all tree-level processes involving $q\bar q$; we do not consider color factors. This implementation has been cross checked both analytically and numerically against independent codes based on {\tt FeynArts}, {\tt FormCalc} and {\tt LoopTools} \cite{Hahn:2001rv}. These tools are also used to numerically compute all $gg$-initiated processes. For axial-vector production at one loop there arises a subtlety, related to nested anomaly diagrams that give rise to residual UV divergencies unless a coupling relation between left and right chiral top and bottom quarks is enforced: $g_{b}^{L}-g_{b}^{R}=g_{t}^{R}-g_{t}^{L}$. Note that this coupling relation is fulfilled by the SM $Z$ boson interaction, which is directly related to the renormalizability (and hence unitarity) in the SM. In the axial vector cases we therefore have to include the bottom quark as a propagating degree of freedom, while we only consider the top quark in the loop in the vectorial case. However, as the tree-level constraints are strongest for the axial-vector model, the loop-level subtleties of the axial-vector model turn out to be unimportant and are not presented here.

The partial wave projection is performed by integrating the matrix element over the scattering angle in the two-to-two processes which only depends on two partonic Mandelstam variables $\hat{s},\hat{t}$, or equivalently $\sqrt{\hat{s}},\cos\theta$, as well as Wigner functions \cite{Jacob:1959at}. In this work we focus on the $s$-wave contribution
\begin{equation}
\label{eq:unitviol}
a_0(\sqrt{\hat{s}}) = {i\over 32 \pi}\int_{-1}^{1}  {\cal{M}}(\sqrt{\hat{s}},\cos\theta)~\hbox{d}\cos \theta\,,
\end{equation}
and we loop over helicities to find the maximally contributing helicity combination instead of calculating the unpolarised matrix element.
 We adopt $|a_0|\geq 1$ as the constraint on the parameter region where unitarity is violated.\footnote{It should be noted that considerations of loop corrections are likely to yield more stringent constraints. This can be seen from the case of the SM where longitudinal $WW$ scattering sets a constraint $m_H \sim  \hbox{TeV}$. However, even for masses below the $WW$ threshold the decay $H\to 4\ell$ is bigger than $H\to 2f$ and since these decay widths are related to the imaginary parts of the Higgs boson self-energy, perturbation theory is challenged at a much lower scale than suggested by investigating tree-level unitarity alone, see \cite{Goria:2011wa}.} In practice, the critical coupling strength at a given centre-of-mass energy can be extracted from a single integration as the matrix elements scale linearly with the coupling strength. We adopt finite light flavor quark masses throughout, in particular to avoid the massless (and hence divergent) $t$ channels.

Perturbative unitarity may thus break down in two ways.  It may be that kinematic quantities in the amplitude grow uncontrollably, signaling that the underlying structure is incomplete and calls for new interactions or fields. Alternatively, even in a theory which is not pathological, if the relevant couplings become very large perturbation theory may break down. In this case it is expected that higher order terms in the perturbative expansion (at higher loop levels) become comparable to the leading order terms and all of these terms will in fact lead to an amplitude which respects perturbative unitarity.  In practice, when a coupling becomes so large as to violate unitarity we must abandon the use of perturbation theory to calculate scattering amplitudes.

\subsection{Results}
The main results of this section are shown in \Fig{fig:TreeLimits}, \Fig{fig:LoopLimits} and \Fig{fig:ZLimits}.

\subsubsection{Tree Level Monojets}
\begin{figure}[t]
\begin{centering}
\includegraphics[width=0.9\textwidth]{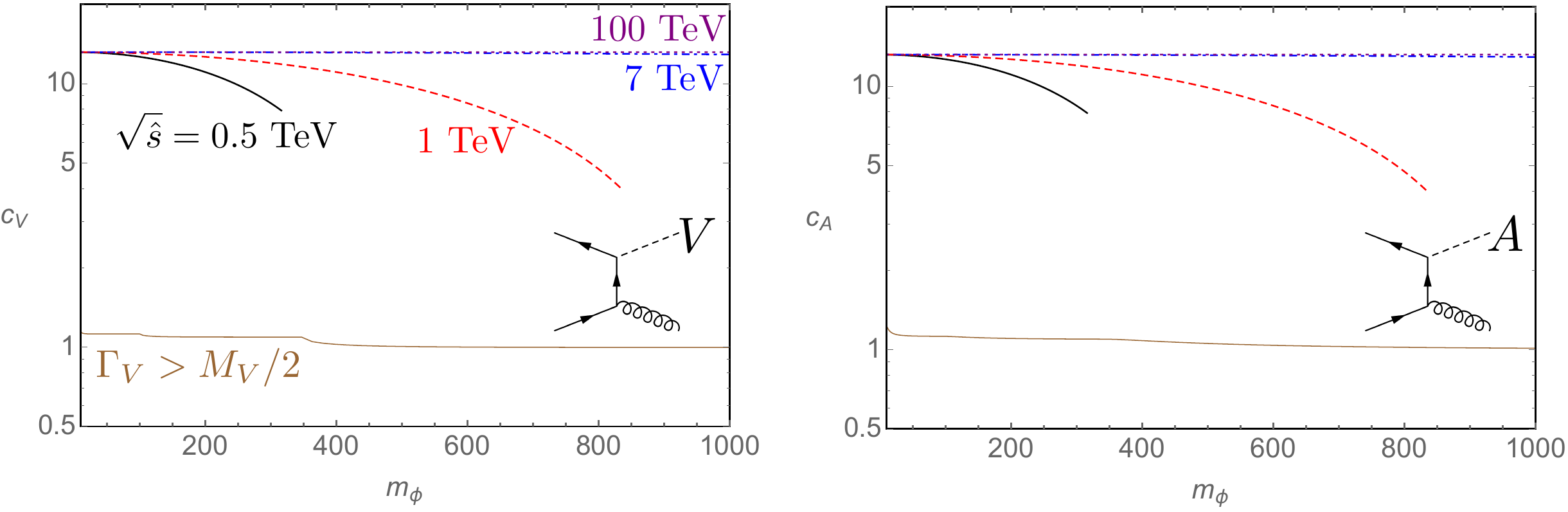}
\caption{Contours of couplings which violate perturbative unitarity at various fixed CM energies for varying mediator mass $m_\phi$.  Vector (V) and axial-vector (A) mediators are shown.  Contours of fixed width-to-mass ratio are also shown.}
\label{fig:TreeLimits}
\end{centering}
\end{figure}

Let us first discuss \Fig{fig:TreeLimits}.  In this figure the tree-level monojet signature ($\overline{q} q \to g + \phi$, $\phi \equiv V,A$) is considered.  Scalar and Pseudoscalar mediators are not shown as their couplings to light quarks are typically taken to be proportional to the Higgs Yukawas, in which case the dominant production mechanism is at one-loop.  This is discussed in the next section.

In \Fig{fig:TreeLimits} contours for partonic center of mass energies $\sqrt{\hat{s}}= 0.5,1,7,100$ TeV are shown.  We choose partonic CM energies of $7$ and $100$ TeV because the former is representative of a very high energy event at the LHC, and the latter of the highest partonic energy that could be imagined in a future collider.  In practise, we will see that already $\sqrt{\hat{s}}= 1$ TeV is sufficient to understand the (un)importance of the unitarity bounds on mediator production.   Each contour describes the simplified model coupling at which perturbative unitarity breaks down according to the prescription of \Sec{sec:unit}.  In addition, for each model a contour shows the coupling value at which the mediator width becomes as large as half the mediator mass, in which case a simple particle description begins to break down.

For each mediator scenario it is clear that as $\sqrt{\hat{s}}$ is increased, the coupling which will lead to unitarity violation becomes smaller, however this effect is negligible between $\sqrt{\hat{s}}= 7$ TeV and $\sqrt{\hat{s}}= 100$ TeV for the axial vector case.  Interestingly, for all mediators the coupling which leads to unitarity violation is approximately the same for the larger CM energies $\sqrt{\hat{s}} = 7, 100$ TeV. This results from the fact that the amplitude has lost its memory of the mediator mass when evaluated at much larger energies and also the fact that the amplitude is well-behaved as all potentially dangerous contributions are suppressed by the small quark mass.  Another interesting feature in each plot is that as $\sqrt{\hat{s}} \to m_\phi$ the coupling at which unitarity is violated becomes smaller.  The reason for this is however artificial, since the final state gluon is becoming soft in this limit and the IR divergence is manifesting as a large logarithm which significantly increases the cross section.  Inclusion of the NLO QCD corrections to $\overline{q} q \to \phi$ in this case would regulate this unphysical divergence.  For this reason we truncate the contours early at $m_\phi= \sqrt{\hat{s}} \sqrt{1- 2 E_j/\sqrt{\hat{s}}}$ where we take $E_j = 150 \text{ GeV}$.  The motivation for this choice is that as the final-state jet energy falls below $150$ GeV the jet $p_T$ would fall below typical cuts in any case and these events would instead contribute to the inclusive mediator production cross section instead.

For both mediators we see that perturbative unitarity does not become a relevant constraint unless the couplings are large, unsurprisingly around $c\sim \mathcal{O}(4 \pi)$.  Also, the couplings which do violate unitarity are typically so large that the particle interpretation of the mediator is called into question by the large mediator width.  This result is not surprising for the vector mediator model as it is in principle already UV-complete\footnote{For example, for a vector mediator model one could imagine a gauged $\U(1)_{B-L}$ symmetry with a Stuckelberg mechanism providing the longitudinal degree of freedom for the massive gauge boson.} and thus we do not expect to see perturbative unitarity violation unless the relevant couplings have themselves become nonperturbative.   For the axial vector case it was shown in \cite{Kahlhoefer:2015bea} that in the absence of additional Higgs-like fields processes such as $\overline{q} q \to A^\ast \to \overline{q} q$ may violate perturbative unitarity whenever $m_{A} \ll m_q$.  This is essentially  because for a gauge coupling $g_A \sim \mathcal{O}(\text{1})$, in the limit $m_{A} \ll m_q$ an axial-vector interaction essentially implies a large non-perturbative coupling between the longitudinal (Goldstone) component of the massive axial vector and the quarks.  However, in the case considered here, since production is from initial state light quarks, such effects are highly suppressed by powers of the small quark Yukawas.\footnote{We have confirmed that whenever the initial state quark masses are large, as for e.g. top quarks, this effect does become relevant and lead to a breakdown of perturbative unitarity for the axial vector case.  However, this is not relevant for colliders as the light quarks dominate the production.}  Thus although potential issues with unitarity may arise, they are so small that the picture for the axial vector is qualitatively similar to the vector case.

To summarize, for the tree-level monojet processes we see that at the LHC and potential future colliders the vector and axial vector mediator models are unlikely to suffer from problems relating to perturbative unitarity in the DM production processes considered, unless large couplings are invoked.

\subsubsection{Loop Level Monojets}
\begin{figure}[t]
\begin{centering}
\includegraphics[width=0.9\textwidth]{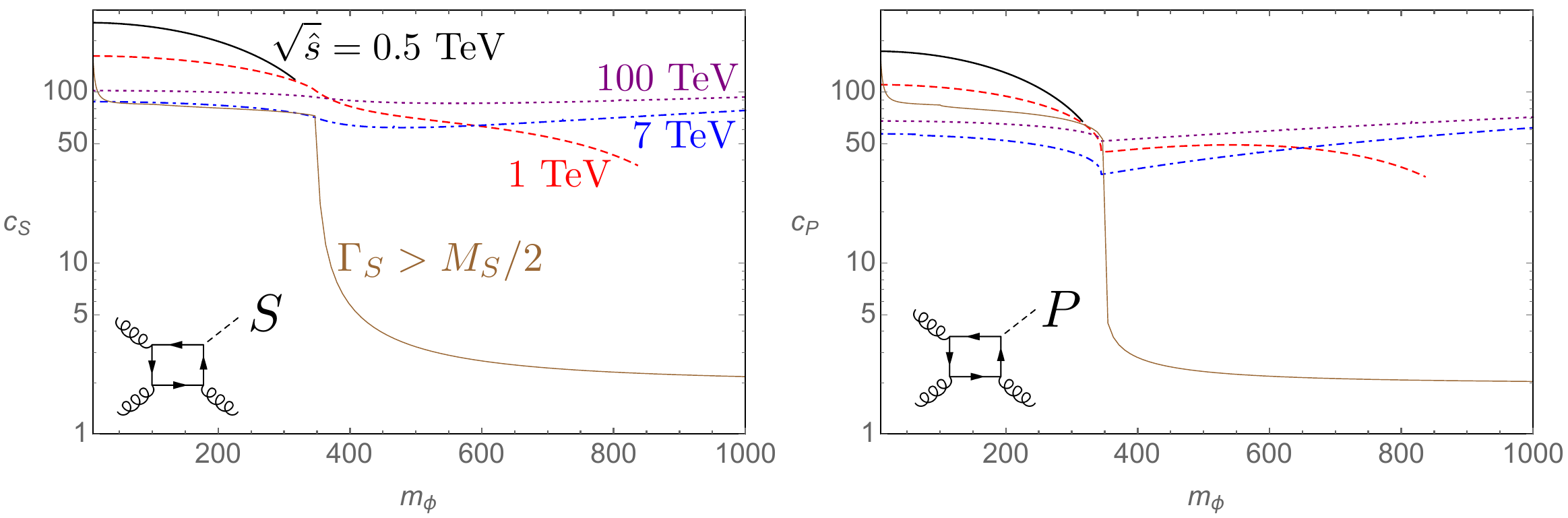}
\caption{As in \Fig{fig:TreeLimits} however for the one-loop gluon-induced monojet process.}
\label{fig:LoopLimits}
\end{centering}
\end{figure}

Let us now consider the one-loop gluon induced monojet processes shown in \Fig{fig:LoopLimits}, relevant for scalar and pseudoscalar mediators.  In this case, due to the additional loop factor and suppression by $\alpha_S$ in the matrix element relative to the tree-level case, perturbative unitarity is satisfied up to much larger couplings, unsurprisingly around $c\sim \mathcal{O}(16 \pi^2)$.\footnote{It has been emphasised to us by Felix Kahlhoefer that associated production of the mediator with top quark pairs may also lead to interesting bounds, although a consideration of perturbative unitarity in $2\to3$ processes would be required.}

Keeping in mind the fact that $c_S$ and $c_P$ are the factors rescaling the top quark Yukawas, for scalar and pseudoscalar mediators perturbative unitarity is preserved up to couplings so large that they would be considered non-perturbative in any case.  At very high energies, for mediator masses $m_\phi < 2 m_t$ the couplings which lead to violation of perturbative unitarity are smaller than those that lead to a large mediator width for the pseudoscalar mediator.  Interestingly, we do see a difference between the limits at $\sqrt{\hat{s}} = 7, 100$ TeV, however since the loop factor is small this is irrelevant in any case.

To conclude, in gluon-initiated one-loop monojet processes perturbative unitarity is unlikely to become an important consideration in parameter regions considered for dark matter processes unless very large couplings are considered.

\subsubsection{Mono-$Z$}
The mono-boson constraints are perhaps most interesting as they directly probe the electroweak symmetry breaking structure of the SM and hence at high energies are sensitive to the gauge structure of the simplified model couplings.  Here we will consider mono-Z signatures, which are similar to the mono-W signatures in respect of perturbative unitarity.  Once again, we only consider the vector and axial vector mediators as the scalar and pseudoscalar couplings to light quarks are suppressed by the small Yukawas.

\begin{figure}[t]
\begin{centering}
\includegraphics[width=0.9\textwidth]{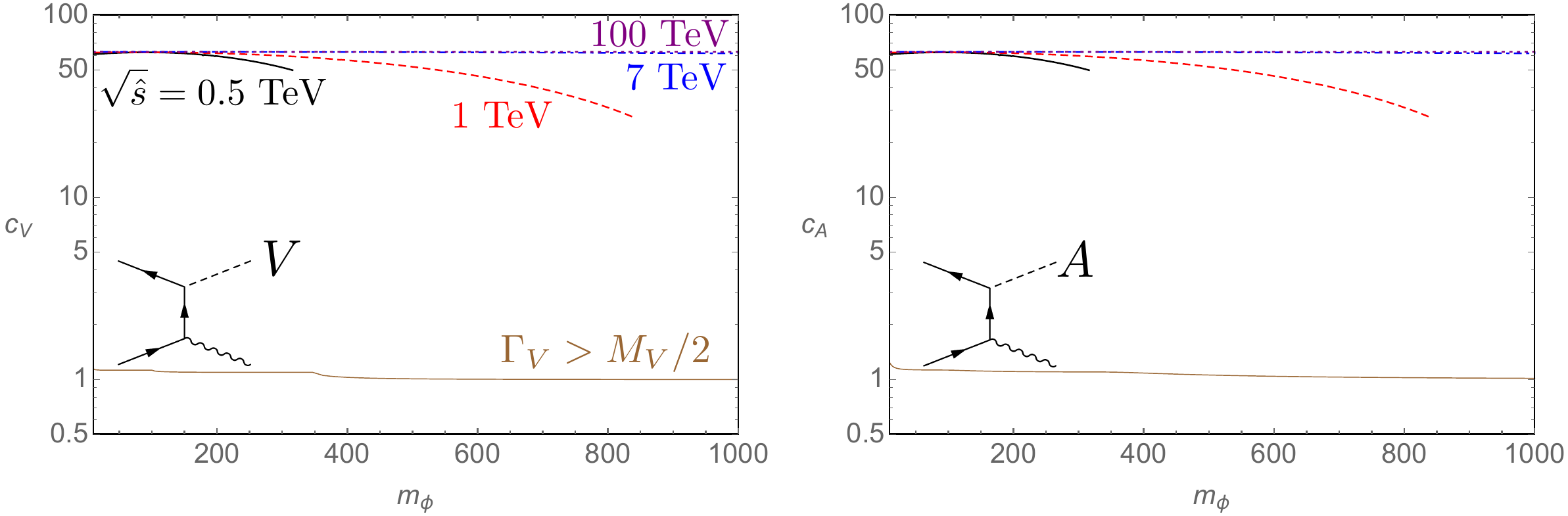}
\caption{As in \Fig{fig:TreeLimits} however in this case mono-Z signatures are considered.}
\label{fig:ZLimits}
\end{centering}
\end{figure}

The early work of \cite{Gunion:1990kf} suggests that in the absence of a new Higgs boson coupled to $Z^\mu A_\mu$ the axial-vector Feynman diagram of \Fig{fig:ZLimits} should violate perturbative unitarity at high CM energies.  However, the dangerous contribution to the amplitude is proportional to the fermion mass, thus we do not expect this to be an important concern.  As shown in \Fig{fig:ZLimits}, the couplings for which perturbative unitarity breaks down are all very large, showing once again that unitarity is only violated as a result of the breakdown of perturbativity. Furthermore, as long as the mediator mass is less than half of its width, perturbative unitarity is satisfied.
\begin{wrapfigure}[13]{r}{8cm}
  \centering
  \includegraphics[width=0.5\textwidth]{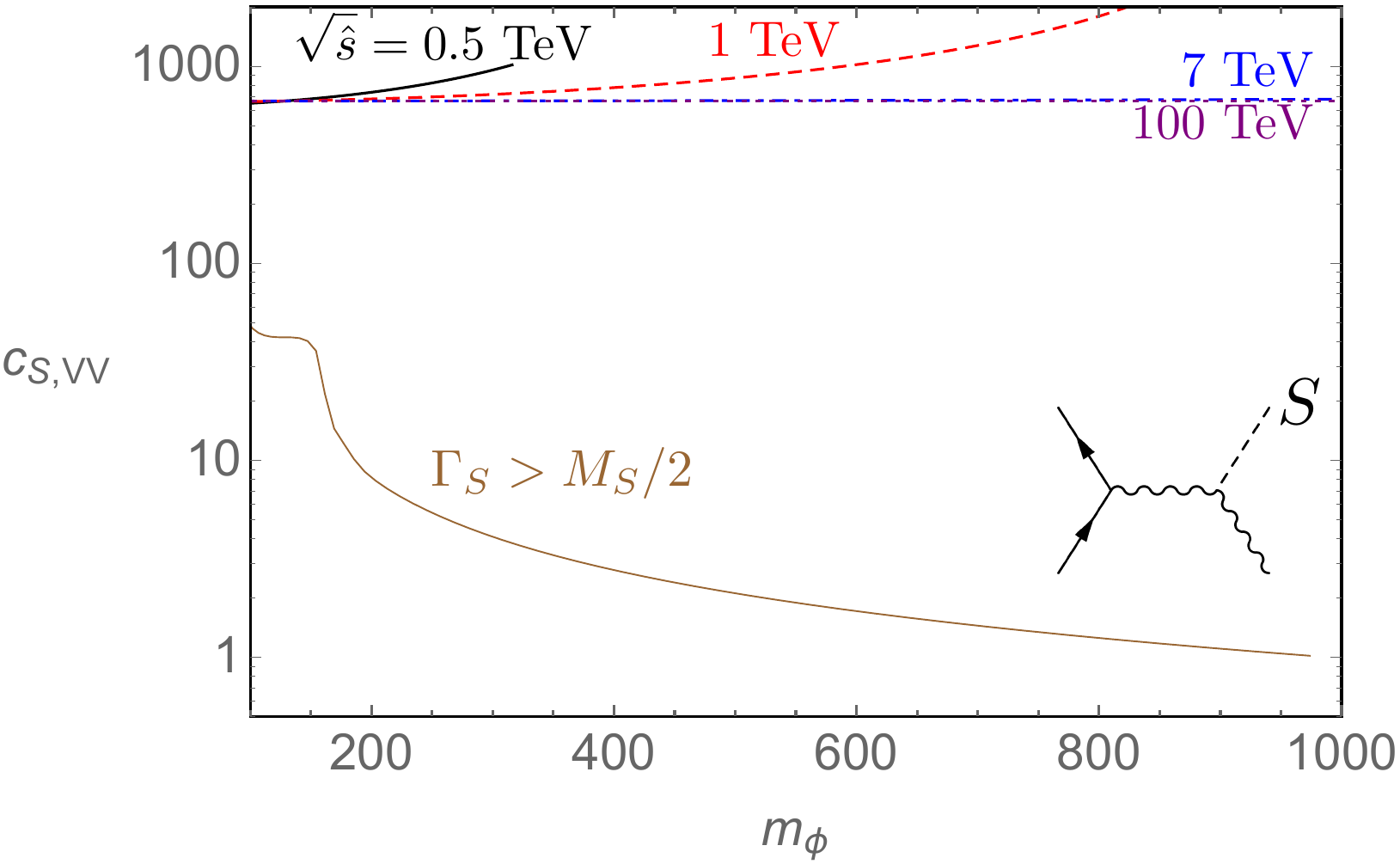}
  \caption{As in \Fig{fig:TreeLimits} however in this case mono-Z signatures are considered.}
  \label{fig:NewLimits}
\end{wrapfigure}

\subsubsection{Fermiophobic Mediator}
Finally we consider the new fermiophobic scalar mediator simplified model proposed in \Sec{sec:newmodel}.  This model only contains a scalar mediator.  We show the perturbative unitarity constraints in \Fig{fig:NewLimits}, where we have considered the mono-Z process.  The coupling $c_{S,VV}$ rescales the Higgs coupling, which is determined by the weak couplings, thus a weak coupling enters in the fermion coupling to the Z-boson.  Due to this suppression, perturbative unitarity is preserved up to very large values of $c_{S,VV}$ and it is likely that the width constraint is violated before perturbative unitarity becomes an issue.  As before, the work of \cite{Gunion:1990kf} showed that perturbative unitarity will in general be violated if an additional coupling of $S$ to the initial state fermions, proportional to their mass, is not included.  However, as in the axial vector case, for the light quarks this effect is negligible.

\section{Conclusions}\label{sec:conclusions}
As LHC searches for DM resume at 13 TeV, the simplified model approach has emerged as a popular option for studying missing energy signatures.  In this work we have argued that in principle many of the simplified model descriptions will break down at some high energy scale.  This presents an opportunity to consider the precise role that simplified models are expected to play in the search for DM at colliders.  A prime diagnostic for the health of a theoretical description of a process is perturbative unitarity, and this has already emerged as an interesting consideration with respect to simplified DM models \cite{Bell:2015sza,Chala:2015ama,Kahlhoefer:2015bea,Haisch:2016usn}.  We have argued that since simplified models are introduced to model missing energy signatures, it is pragmatic to require that perturbative unitarity is observed in the collider process under question.

We quantified the couplings in popular simplified models that will lead to perturbative unitarity violation in the specific processes used to search for DM at the LHC.  We have found, unsurprisingly, that once a coupling is chosen small enough that the mediator width is less than half the mediator mass, perturbative unitarity constraints will be satisfied.   Moreover, the couplings for which perturbative unitarity breaks down are as large as would be expected from na\i ve estimates for the breakdown of perturbation theory.  Any potential structural issues that may lead to violation of unitarity at high energies are typically suppressed by powers of the incoming quark masses, which are very small for light quark initial states at the LHC.    Hence, the presented analysis encourages to \emph{keep calm and carry on} with the use of simplified models in searches for Dark Matter at the LHC.

We also sketched a broad discussion of the underlying issues that may cause a UV-incomplete simplified model description to break down at high energies.  This was aimed at a non-expert reader.  This discussion also highlights the need to consider a complete set of simplified models in order to effectively capture the full collider phenomenology possible in UV-complete models.  In the process of considering various possible structures in the UV, we have proposed a `Fermiophobic' scalar mediator simplified model as an interesting addition to the models already under consideration.  This model enlarges the coupling structures possible for scalar mediators and also highlights the importance of VBF-motivated missing energy searches alongside the usual mono-object processes.

\vskip 1 \baselineskip

\noindent {\it{Acknowledgments.  We thank P. Fox, G. Giudice, P. Harris, F. Kahlhoefer, Kai Schmidt-Hoberg and R. Torre for enlightening discussions.  MS is supported in part by the European Commission through the ``HiggsTools'' Initial Training Network PITN-GA-2012-316704.}}

  
\bibliographystyle{JHEP}
\bibliography{references}

\end{document}